\documentstyle[pre, aps, epsfig]{revtex}
\begin{document}
%
\title{Dynamical temperature study for classical planar spin systems}
\author{Wira B. Nurdin\footnote{E-mail: nurdin@physik.fu-berlin.de,
on leave from Physics Department, Hasanuddin University, Makassar.},
Klaus-Dieter Schotte\footnote{E-mail: schotte@hmi.de}
\\
Institut f\"ur Theoretische Physik, Freie Universit\"at Berlin\\
Arnimallee 14, 14195 Berlin, Germany
}
\date{\today}
\maketitle

\begin{abstract}
Making use of the Rugh's micro--canonical approach to temperature we study
$XY\!$--spin systems by the strictly energy conserving over-relaxation algorithm.
In this micro-canonical simulation the temperature
and also the specific heat are determined as averages of expressions easy to
implement. The $XY\!$--chain is studied for a test. The second order
transition on a cubic lattice and the first order transition on fcc lattice
are analyzed in greater detail to have a more severe test about the
feasibility of this micro-canonical method.
\end{abstract}
\smallskip
\leftline{\hbox to 0.75in{}PACS numbers: 05.50.+q, 75.10.Hk, 75.40.Mg}

\section{Introduction}
Dynamical simulations of large systems of spins or particles should in
principle lead to the same information as statistical simulations generating
a canonical ensemble. Although the energy is conserved one can determine
the temperature for a system consisting of kinetic and potential
energy the temperature by the average kinetic energy.
With the recent advance in the understanding of the micro--canonical
temperature initiated by Rugh\cite{Ru97,Ru98,Nu00},  one can ``measure'' 
the temperature and also the specific heat along the trajectory of spin systems 
where such a decomposition of the energy does not exist. In the case of
 $XY\!$--spins one could also add the sum over the angular velocities
but a dynamical check of statistical mechanics by a common molecular
dynamics simulation is not an easy numerical task\cite{Firenz2t}.
We think by using spin systems alone one has more efficient means to pursue
this program.

Several attempts have been made to establish a strictly micro--canonical
approach to study phase transitions \cite{Ja98,Hu98}. Here we want to present
a study of the planar or $XY\!$--spin systems, making use of a micro--canonical
temperature definition discussed recently \cite{Nu00} 
and the ``over--relaxation algorithm'' \cite{Bro87,Cre87,Li89}. 
This algorithm has been used to accelerate Monte Carlo simulations 
for Heisenberg or $XY$ spins for example.
Since it conserves energy it can be taken as a Metropolis step,
where steps leading to equal or lower energy are always accepted.
The Metropolis procedure leads to thermalisation, whereas the
over--relaxation algorithm generates different configurations and
thereby improves the convergence.
The question we want to address is, can one work alone with
the over--relaxation algorithm. Since the temperature can be calculated
as an average the Metropolis procedure is not needed for the
temperature determination. However, it is not guaranteed whether in
a fixed energy simulation one ``visits'' enough regions of phase
space in order to determine the physical observables reliably.
More specifically we want to ask: is the finite size analysis the same
as for canonical simulations? A closely connected question: is the
analysis of a first order transition really different from
the standard one using a histogram\cite{Ja98} analysis?

For numerical tests we have chosen $XY\!$--spins on a chain and on three
dimensional simple cubic (sc) and face centered cubic (fcc)
lattices both with nearest neighbour coupling. 
A second order transition on the sc lattice and the 
first order one on fcc\cite{Diep89} will thus be studied.

\section{Micro--canonical temperature for classical $XY\!$--spin systems}
\subsection{General approach}

Instead of three components for the classical Heisenberg system in our
preliminary paper\cite{Nu00}, here we want to study the simpler
$XY$--spin system where only two components
$\vec S = \bigl( S^1,\, S^2 \bigr)$ have to be taken into account,
that is the spin vector is confined to a circle with radius $S$.
No dynamical equation can be written down
similar to $\partial \vec S \big/ \partial t = \vec S \times \vec H$ for
the three dimensional vector $\vec S$ in a magnetic field $\vec H$.
The energy $-\vec H \cdot \vec S = -H\,S \cos \varphi$ is the same
for the two directions $\pm \varphi$ the vector $\vec S$ can point.
Contrary to the three dimensional case, there is no continuous path
between these equal energy states at $\pm\varphi$.

Still we can calculate the size of the phase space $W$ summing over
all contributions with the same energy $E$.
$W$ corresponds to the density of states in the classical limit.
With the entropy $S$ as a function of energy given by
\begin{equation}
S = \ln W
\label{2.5}
\end{equation}
(Boltzmann's constant set to unity), the density of states contains all
information necessary for a thermodynamic analysis.  We want to show
that the derivative of $S$ with respect to the energy is easily accessible.

With $N$ spins of length $|\vec S_i| = 1$, that is
$\vec S_i = \bigl(\cos\varphi_i,\, \sin\varphi_i \bigr)$ one has
\begin{equation}
W \, = \, \int \delta(E - {\cal H})\,
\hbox{$\prod_{i=1}^N$} \, d \varphi_i
\, = \, \int\displaylimits_{{\cal H} = E}
{d O_{N-1} \over |\nabla {\cal H}|}
\label{2.1}
\end{equation}
where the integration is over $N$ unit circles.
However, the integral is restricted to $2N\!-\!1$ dimensions by
${\cal H} = E$. This restriction taken into account, one arrives at
the second form of the integral with the gradient taken with
respect to all $\varphi_i$ angles\cite{Khin}.
Because the direction of the infinitesimal surface elements $d \vec O$
normal to the surface of constant energy coincides with the direction of
the gradient $\nabla {\cal H}$, one arrives at a second form of
the constant energy integral
\begin{equation}
W \, = \, \int\displaylimits_{{\cal H} = E}
{\nabla {\cal H} \>d \vec O_{N-1} \over |\nabla {\cal H}|^2}\
\, = \,\int \Theta (E - {\cal H})
\>\nabla \,\left( {\nabla {\cal H} \over \,|\nabla {\cal H}|^2} 
\right) \> \hbox{$\prod_{i=1}^N$} \, d \varphi_i \ .
\label{2.2}
\end{equation}
Applying Gauss' theorem this integral can be written as an integral
over all phase space with energy lower than $E$, the restriction
given by the Heavyside step function $\Theta$. The quantity to be integrated is
the divergence of a vector field $X$ with the property
that the scalar product of $\nabla {\cal H}$ and 
$X = \nabla {\cal H} /  \,|\nabla {\cal H}|^2$ is unity \cite{Ru98}.
The derivative is then
\begin{equation}
{\partial W\over \partial E} \, = \,\int \delta({\cal H} - E) \;
\nabla\,\left({\nabla {\cal H} \over \,|\nabla {\cal H}|^2} \right) \>
\hbox{$\prod_{i=1}^N$} \, d \varphi_i
\label{2.3}
\end{equation}
a constant energy integral like (\ref{2.1}) for $W$. The ratio of 
these two integrals gives then the inverse temperature
\begin{equation}
{1 \over T} \, = \, {\partial W\big/ \partial E \over W}
\label{2.4}
\end{equation}
since $1\big/ T = \partial S \big/ \partial E$
with the entropy $S$ given by (\ref{2.5}).

In order to obtain the inverse temperature according to (\ref{2.3})
and (\ref{2.4}) one has to average the following quantity,
called $1/{\cal T}_{xy}$ for short
\begin{equation}
{1 \over {\cal T}_{xy}} \, = \, \sum_{i=1}^N \partial_i
\> { \partial_i {\cal H}
\over 
\sum_l ( \partial_l {\cal H})^2 } \ , 
\label{2.6}
\end{equation}
where we use the notation $\partial_i = \partial  / \partial \varphi_i$. To
compare with the Heisenberg case\cite{Nu00} of a three dimensional vector
confined to a sphere one identifies $\partial \big/\partial \varphi_i$
as the third component of the angular momentum operator
\begin{equation}
{ \partial \over \partial \varphi_i}
\, = \, S_i^1  {\partial \over \partial S_i^2} \, - \,
S_i^2  {\partial \over \partial S_i^1}
\label{2.7}
\end{equation}
used in quantum mechanics, only ``$i$'' is missing. Eq.\thinspace\ref{2.6}
can then be viewed as a reduced form of a micro-canonical temperature for
a Heisenberg spin system with the complete angular momentum
operator  $\vec {\cal L}_i = \vec S_i \times \nabla_i$
\begin{equation}
{1 \over {\cal T}_{xyz}} \, = \, \sum_{i=1}^N
\vec {\cal L}_i \> {\vec {\cal L}_i {\cal H}
\over \sum_l \bigl( \vec {\cal L}_l {\cal H} \bigr)^2 }
\label{2.8}
\end{equation}
instead of only the third component of $\vec {\cal L}_i$
appearing in eq.(\ref{2.6}).

For large system sizes $N \gg 1$ the quantity to be averaged in (\ref{2.6})
can be simplified. The two terms generated by differentiation are
\begin{equation}
{1 \over {\cal T}_{xy}} \> = \> {\sum_{i} \partial_i^2 {\cal H}
\over
\sum_l ( \partial_l {\cal H})^2 }
\> + \> \sum_{i=1}^N
\partial_i {\cal H} \>
\partial_i { 1 \over \sum_l (\partial_l {\cal H})^2 } \ ,
\label{2.9}
\end{equation}
where the last one can be neglected since it is only of order $1/N$.
The reason is the short range of exchange interactions.
The Hamiltonian $\cal H$ connects a spin $\vec S_i$ only to its neighbors
so that the differentiation with respect to $\varphi_i$ is
nonzero for $l = i$ and $l$ close to $i$. As a result the second term scales
as $\propto 1/N$ since a single sum in the numerator is
outweighed by a double sum in the denominator.
Also for large spin systems neglecting again $1/N$--corrections
one can take the inverse of (\ref{2.9}) to determine the temperature
$T$ directly as an average of
\begin{equation}
{\cal T}_{xy} \> \approx \>
{\sum_i ( \partial_i {\cal H})^2
\over \sum_l \partial_l^2 {\cal H} } \ ,
\label{2.10}
\end{equation}
that is $T = \langle {\cal T}_{xy} \rangle$.

\subsection{Micro-canonical temperature in simulations}
%
Although there exist no dynamical equations which
conserve the energy for  $XY\!$--spin systems, there is a so--called
``over--relaxation algorithm'' in use to generate spin configurations
of equal energy. This algorithm is used in conjunction with the standard
Metropolis procedure, see for example \cite{Li89}, to diminish the
effects of critical slowing down\cite{Cre87}. 
The Hamiltonian we want to study is
\begin{equation}
{\cal H}_{ex} = -\sum_{ij} J_{ij}\cos(\varphi_i - \varphi_j) \ ,
\label{2.11}
\end{equation}
where the sum is over the exchange constants $J_{ij}$ connecting close
spins at sites $i$ and $j$. In an equivalent notation this is
${\cal H}_{ex} = -\sum_{ij} J_{ij}\, \vec S_i \cdot \vec S_j$ with
$\vec S_i = (\cos \varphi_i, \, \sin \varphi_i)$. The energy
$E_i$ contributed by an individual spin $S_i$ is
\begin{equation}
E_i = -\vec S_i\cdot \hbox{$\sum_{j}$} J_{ij}\,\vec S_j =
-\vec S_i\cdot \vec H_i
\label{2.12}
\end{equation}
with ${\cal H}_{ex} = \hbox{$1\over 2$} \sum_i E_i \,$.
To get the total energy ${\cal H}_{ex}$ one has to sum over all sites
of the lattice and since the exchange interaction appears twice
the sum has to be divided by two. The ``local'' energy $E_i$ depends on the
``molecular'' field $\vec H_i$ generated by the neighbors. It does not
change if
\begin{equation}
\vec S_i \> \Longrightarrow \>
-\vec S_i \, + \, 2\,\vec H_i \bigl(\vec S_i\!\cdot\!\vec H_i\bigr) /H_i^2
\ .\label{2.13}
\end{equation}
This is the basic step of the ``over--relaxation'' algorithm which in
geometrical terms means a change of $\vec S_i$ to its mirror position
with the mirror line given by $\vec H_i$ as in Fig.\thinspace 1. In the
simulation the spins to be updated are mostly chosen randomly\cite{Cruz}.

Using the special form for the Hamiltonian (\ref{2.11}) the formula
for the temperature (\ref{2.10}) simplifies to
\begin{equation}
{\cal T}_{xy} \> = \> -
{\sum_i ( \partial_i {\cal H}_{ex})^2
\over 2\,{\cal H}_{ex} } \ ,
\label{2.14}
\end{equation}
since summing the second derivative of the Hamiltonian ${\cal H}_{ex}$
with respect to all angles $\varphi_i$ one recovers ${\cal H}_{ex}$
merely multiplied by $-2$.
The easiest way to test the validity of the temperature determination by
averaging (\ref{2.14}) is to apply the over--relaxation algorithm to
a spin chain. However, this algorithm for a linear chain with
\begin{equation}
{\cal H}_{chain} = -J\,\hbox{$\sum_{i=1}^N$} \cos(\varphi_i -
\varphi_{i+1})
\quad\hbox{and}\quad \varphi_{N+1} = \varphi_1
\label{2.15}
\end{equation}
would be non ergodic in the case that all exchange constants are
equal\cite{Dhar}. In alternating the exchange by $\pm 10\%$ along
the chain one has a remedy for this defect easy to implement.
The temperature according to (\ref{2.14}) for a homogeneous chain is
\begin{equation}
{\cal T}_{chain} = - \hbox{$1\over 2$}\hbox{$\sum_{i=1}^N$}
\left[J\,\hbox{$\sum_{\pm}$} \sin(\varphi_i - \varphi_{i\pm 1}) \right]^2
\big/ {\cal H}_{chain} \ .
\label{2.16}
\end{equation}
The task is then to check whether for a large chain the canonical
relation\cite{Stan} between energy $E$ and temperature
\begin{equation}
\epsilon = E\big/N  = - J\,I_1(\hbox{$J/T$}) \big/ I_0(\hbox{$J/T$})
\label{2.17}
\end{equation}
is reproduced. $I_0$ and $I_1$ are modified Bessel function.
The specific heat as the derivative $C = dE/dT$ then is
\begin{equation}
C/N =
(J^2 + T\,\epsilon - \epsilon^2 )/T^2 \ .
\label{2.18}
\end{equation}
We checked that the slight change of the exchange constant
$J = 1.0 \pm 0.1$ had no visible effect in Fig.\ 2.
For a comparison using the average value $J = 1$ is sufficient.
The temperature and the specific heat for a $XY\!$--chain
determined by the over--relaxation method by varying the energy
agree well, as can be seen in Fig.\ 2, where energy and
specific heat are plotted in the conventional way as functions
of the temperature. The question how to calculate the specific
micro--canonically for $XY$--spin system will be addressed in the
next section.

An initial state with a prescribed energy to start the simulation
is found in the following way.
One solves numerically the differential equations
\begin{equation}
\dot \varphi_i = \pm \partial_i\,{\cal H}_{ex}/\tau \ .
\label{2.19}
\end{equation}
A change of the angles $\varphi_i$ of the spins involves of
a change of the total energy, that is $\dot {\cal H}_{ex} = 
\pm(1/\tau)\sum_i (\partial_i\,{\cal H}_{ex})^2$.
Depending on the size and sign of the relaxation time $\tau$
in principle starting from randomly chosen orientations $\varphi_i$ of
spins any prescribed energy could be reached. Very low
energies states, however, can only be found by a succession of
energy conserving over--relaxation steps and steps lowering
the energy by small amounts using (\ref{2.19}).

\section{Specific heat of a $XY$ spin system}
\subsection{Specific heat for the $XY$--chain}
In taking the derivative of the entropy $S$ with respect to the
energy $E$ one obtains the inverse temperature.
The second derivative with the respect to the energy will then determine
the inverse of the the specific heat. Formally the first derivative
defines the inverse temperature by the divergence of a vector $\vec X$
\begin{equation}
\vec X = \Bigl(
\partial_1 {\cal H},\ \partial_2 {\cal H},\ \cdots
\partial_N {\cal H}\Bigr)
\Big/
\sum_{i=1}^N \,\bigl( \partial_i {\cal H} \bigr)^2
\label{3.1}
\end{equation}
according to (\ref{2.6}) written as $1/{\cal T} = \sum_i \partial_i X_i$.
This standard form of $\vec X$ can be changed.
One can modify $\vec X$ as long as the scalar product between
$\vec X$ and $\nabla {\cal H}$ does not change, that is
\begin{equation}
\sum_{i=1}^N X_i \, 
\partial_i {\cal H}
 \> = \> 1 \ .
\label{3.2}
\end{equation}
To get the volume of phase space with (\ref{2.2}) correctly, (\ref{3.2}) is the only
condition\cite{Ru98}. In a ``hydrodynamic'' picture this means
that the flow in phase space $\vec X$ across ${\cal H} = E$
should be the same as the ``natural'' flow given by (\ref{3.1})
which is perpendicular to the ${\cal H} = E$ surface.
The constraint (\ref{3.2}) leaves enough room for choices better fitted
to specific heat calculations as will be shown.

For a chain a variant to $\vec X$ is
\begin{equation}
\vec X' = \Bigl(0,\  \partial_2 {\cal H},\ 0,
\  \partial_4 {\cal H},\ \cdots \>0,\  \partial_N {\cal H}
\Bigr) \Big/
\sum_{i=1}^{N/2} \,\Bigl( \partial_{2i} {\cal H} \Bigr)^2\ ,
\label{3.3}
\end{equation}
and one easily checks that the normalisation (\ref{3.2}) is still valid.
In the following it is assumed that $\cal H$ is of the exchange type
restricted to nearest neighbors like (\ref{2.11}) or specifically
(\ref{2.15}) for the one dimensional case. The inverse temperature
is then given by an expression similar to (\ref{2.9})
\begin{equation}
{1\over {\cal T}'} \, = \,
\sum_{i=1}^{N/2} \partial_{2i} \, { \partial_{2i} {\cal H}
\over
\sum_j \bigl( \partial_{2j} {\cal H} \bigr)^2}
\> = \>
{\sum_{i=1}^{N/2} \partial_{2i}^2 {\cal H}
\over
\sum_j \bigl( \partial_{2j} {\cal H} \bigr)^2}
\> - \> 2 \,
{\sum_{i=1}^{N/2} \bigl( \partial_{2i} {\cal H} \bigr)^2
\partial_{2i}^2 {\cal H}
\over
\bigl[\,\sum_j \bigl( \partial_{2j} {\cal H} \bigr)^2 \bigl]^2}
\label{3.4}
\end{equation}
where in the summations only the even sites along the chain appear.
The second term, although $\propto 1\big/ N$, contributes to the
specific heat as will be seen. Its simpler form compared to the
one derivable from the original expression (\ref{2.9}) for $1/{\cal T}$
is helpful for an easier numerical calculation.

For the temperature determination the inverse of (\ref{3.4}) without the
$1/N$--correction is sufficient, that is we take
\begin{equation}
{\cal T} \, = \,- \sum_{j=1}^{N/2}
\Bigl( \partial_{2j} {\cal H} \Bigr)^2
\Big/ {\cal H}
\label{3.5}
\end{equation}
and determine the temperature by $T = \langle {\cal T} \rangle$. The energy
$\cal H$ or $E$ in the denominator is obtained by summing up the second
derivatives, that is $\sum_{i=1}^{N/2} \partial_{2i}^2 {\cal H} =
-{\cal H}\,$. The derivative of $T$ with respect to temperature consists
of two contributions
\begin{eqnarray}
{d T\over d E} \, = \, {1 \over C} \, &=& \,
\left\langle\,
\sum_{i=1}^{N/2} \partial_{2i} \,
\frac{ \partial_{2i}{\cal H} }{ \sum_j \bigl(\partial_{2j} {\cal H}\bigr)^2 }
\> {\cal T} \right\rangle
\, - \, \Bigl\langle {\cal T} \Bigr\rangle
\left\langle {1\over {\cal T}'} \right\rangle 
\nonumber \\
&=& \>
1 - {T\over E} \> - \> T  \left\langle {1\over {\cal T}} \right\rangle
\> + \> 2 \,T\,\left\langle
{\sum_{i=1}^{N/2} \bigl( \partial_{2i} {\cal H} \bigr)^2
\partial_{2i}^2 {\cal H}
\over
\bigl[\,\sum_j \bigl( \partial_{2j} {\cal H} \bigr)^2 \bigl]^2}
\right\rangle \ .
\label{3.6}
\end{eqnarray}
The first has its origin in the direct energy dependence of the temperature
average of $\sum_i \partial_i (X_i' {\cal T})$. The second
term comes from the energy dependence of the normalization necessary for the
temperature average. The definition (\ref{3.5}) leads to the first
two terms in the expanded form of the next line whereas the two last
terms are generated by using (\ref{3.4}) for $1/{\cal T}'$. With
\begin{equation}
{1\over {\cal T}} \> = \> {1\over T + \Delta{\cal T}} \> = \>
{1\over T} \> - \> {\Delta{\cal T}\over T^2} \> + \>
{(\Delta{\cal T})^2\over T^3} - \ \cdots
\label{3.7}
\end{equation}
and $\Delta{\cal T} = {\cal T} - T$ the `1' in (\ref{3.6}) disappears and
only terms $\propto 1/N$ remain. This is to be expected since the
reciprocal of the specific heat behaves as $dT/dE \propto 1/N$.
One notices that the inverse of the specific heat depends on
the temperature fluctuation. As a further simplification the averages
of the last term of (\ref{3.7}) can be taken separately for the
denominator and the numerator by neglecting $1/N^2$ corrections.
Further the average of the denominator is simply $(E\,T)^2$ since
$ \bigl\langle \sum_j \bigl( \partial_{2j} {\cal H} \bigr)^2 \bigr\rangle
= T\,E$ according to (\ref{3.5}). The difference to the true
average would result again in a $1/N^2$ correction and can therefore
safely be neglected.

Taking the simplifications into account, the specific heat is then
\begin{equation}
{1 \over C} \, = \,- {T\over E}
\, - \,
{\bigl\langle(\Delta{\cal T})^2 \bigr\rangle\over T^2}
\> + \> 2 \,
{\bigl\langle \sum_{i=1}^{N/2} \bigl( \partial_{2i} {\cal H} \bigr)^2
\partial_{2i}^2 {\cal H}\bigr\rangle
\over 
T\,E^2 } 
\label{3.8}
\end{equation}
No further simplifications are possible. In the last term appears
the local energy $ -\partial_{2i}^2 {\cal H} $
besides the square of the ``torque'' $ (\partial_{2i} {\cal H})^2 $
used for the temperature determination.
To get an understanding of the average  of these two quantities
let us define a local energy deviation $\Delta{\cal H}_{2i} =
- \partial_{2i}^2 {\cal H} - 2\,\gamma\,{\cal H}/ N$.
The $\gamma$ value is determined by splitting the average of
(\ref{3.8}) into two parts
\begin{equation}
\left\langle \hbox{$\sum_{j=1}^{N/2}$} \bigl(
\partial_{2j} {\cal H} \bigr)^2 \partial_{2j}^2 {\cal H}
\right\rangle
= - \left\langle \hbox{$\sum_{j=1}^{N/2}$}
\bigl( \partial_{2j} {\cal H} \bigr)^2 \Delta{\cal H}_{2j}
\right\rangle
 - {2\,\gamma \over N}
\left\langle
\hbox{$\sum_j$} \bigl( \partial_{2j} {\cal H} \bigr)^2 \,{\cal H}
\right\rangle
\, = \,{2\,\gamma\over N}\>T\,E^2 \ .
\label{3.9}
\end{equation}
such that the first term depending on the local energy fluctuations
should vanish. The average of the second term is $\propto T\,E^2$
according to (\ref{3.5}). The inverse of the specific
heat (\ref{3.8}) is then
\begin{equation}
{1 \over C} \, = \,- {T\over E} \, + \, {4\,\gamma \over N}
       \, - \,
{\bigl\langle(\Delta{\cal T})^2 \bigr\rangle\over T^2}
\end{equation}
which at low temperatures with $T/(-E) \ll 1$ should reduce to
\begin{equation}
{1 \over C} \, \approx \, {4 \over N}
 \, - \,
{\bigl\langle(\Delta{\cal T})^2 \bigr\rangle\over T^2} \ \>,
\label{3.10}
\end{equation}
a form derivable (see appendix) for a chain of harmonic oscillators with
the temperature fluctuation $2/N$ so that $C_{osc} = N/2$. At
low temperatures one expects $\gamma =1$, since the $XY\!$--spin chain
should have the same specific heat as a harmonic oscillator chain.

To get the specific heat of the $XY\!$--chain shown in Fig.\thinspace 2
relation (\ref{3.8}) has been used. At low temperatures the specific heat
for one spin is $1/2$ as expected.

\subsection{Specific heat for the $XY$--lattice systems}
The strategy to calculate the specific heat for a chain can also
be used for two or three dimensional lattices. For square and simple
cubic lattices a decomposition in two sublattices analogously to the even
odd decomposition for the chain can be made. A spin is even or odd if
the sum of its coordinates $(i,\,j,\,k)$ is even or odd.

The point is that the no nearest neighbor ``bond''
$\vec S_i\! \cdot \!\vec S_j$ appears twice in different components
which is the case for the vector $\vec X'$ defined by (\ref{3.3}).
As a consequence only local energy and temperature terms
have to be averaged.
For a chain and a simple cubic lattice starting from all ``even''
or all ``odd'' sites all the interactions $\vec S_i\! \cdot \!\vec S_j$ are
taken into account and $\sum_{i=1}^{N/2} \partial_{i,l}{\cal H} =
-{\cal H}$ for $l=$ {\sl even}, {\sl odd}.
So after renumbering the spins (\ref{3.8}) can be used without further
changes. As an example we show in Fig.\thinspace 3 the specific heat of the
$XY$--model on a simple cubic lattice obtained this way. Only close to the
critical region differences between the canonical simulation using
the Metropolis method and the micro--canonical simulation are visible.

In the general case, for a face centered cubic lattice as an example,
slight changes have to be made.
The fcc lattices can be thought to consist of four sublattices,
one simple cubic lattice and three additional ones shifted in the three
direction $(1,\,1,\,0)$, $(1,\,0,\,1)$ and $(0,\,1,\,1)$ to the centers
of the faces of a cube. If one picks out one of these simple cubic
lattices one can take over all the equations from the last section
for the chain, only the energy connected to that sublattice is no longer
a constant since is not the total energy. With $M^3$
of lattice points of the simple cubic lattice, that is $N =4\,M^3$
the total number the spins, the temperature analogous to (\ref{3.5}) is
\begin{equation}
{\cal T}_l \, = \,- \sum_{i=1}^{N/4}
\Bigl( \partial_{i,l} {\cal H} \Bigr)^2 \Big/ {\cal H}_l
\quad\hbox{with}\quad
T = \langle {\cal T}_l \rangle  \ ,
\label{3.11}
\end{equation}
where the summation takes into account all angles $\varphi_{j,l}$ of one of
the cubic sublattices with number $l = 1,\,2,\,3\>{\rm or}\,4$. The partial
energy is ${\cal H}_l = -\sum_{i=1}^N \partial_{i,l}^{\,2} {\cal H} \,$. In a
sense one has four thermometers which should of course give
the same temperature.
Similarly one can calculate the specific heat starting from one
sublattice. We studied the anti-ferromagnetic transition on a fcc
lattice (see section V), but the specific heat calculation with
formulas similar to (\ref{3.8}) did not give stable results in the
neighborhood of the first order transition because of the singular
nature of the specific heat.

\section{Scaling for the micro-canonical temperature}
In principle the scaling behavior close to a second order phase
transition should be the same for a canonical ensemble and for a
micro-canonical ensemble. So one expects for a micro-canonical ensemble
close to the critical energy that the ``measured'' temperature should
dependent on the size of the system. This corresponds to the size
dependence of the energy for a canonical ensemble at the critical
temperature\cite{Barber}.

The correlation length $\xi$ grows like
$\xi \propto \xi_0/|T - T_c|^{\,\nu}$ for a temperature close to
the critical temperature $T_c$. This length $\xi$
cannot become larger than the size $L$ of a finite system.
In a micro-canonical simulation at the critical energy $E_c$
this finite size effect should be directly visible. Since the
system size takes the role of the correlation length, the recorded
temperature $T_L$ should deviate from its critical value $T_c$
according to
\begin{equation}
T_L = T_c\, + \,T_1\,L^{-1/\nu} \ ,
\label{4.1}
\end{equation}
that is by rewriting $|T_L - T_c|\, \propto \,L^{-1/\nu}$ as an equation.
In Fig.\thinspace 4 the size dependence of the critical temperature is
plotted and one sees that the temperature defined as an
average of (\ref{2.6}) reproduces the expected effect.

We took as the critical temperature $E_c = 0.989$ found by
Schultka et al.\thinspace\cite{Schu95}. In the temperature scaling
analysis we should recover the temperature these authors started with
that is the value $T_c = 2.2017$ determined by Janke\cite{Ja90}.
This is actually the case, since we find $T_c = 2.202$.

For a canonical system where the temperature is fixed by its critical
value $T_c$ the critical energy $E_c$ has finite size corrections.
Loosely speaking one has
$|E - E_c| \propto |T - T_c|^{1 - \alpha}$ in the neighborhood
of the critical point. The linear dependence between energy changes
$E - E_c$ and temperature changes $T - T_c$ is modified to a power law
dependence by the specific heat exponent $\alpha$. With this modification
the finite size correction of the critical energy follows
in close analogy to (\ref{4.1}) 
\begin{equation}
E_L = E_c + E_1\,L^{-(1-\alpha)/\nu}
\label{4.2}
\end{equation}
from $|E_L - E_c| \propto L^{-(1-\alpha)/\nu}$.
The relation (\ref{4.2}) has been used to determine the critical
energy\cite{Schu95}.
For an $XY$--system practically the same exponent for the micro-canonical
scaling according to (\ref{4.1}) or the canonical one using (\ref{4.2})
can be taken, since $\alpha$ is very small\cite{Lipa}. Actually in
(\ref{4.2}) a correction term $E_2\,L^{-2\,(1-\alpha)/\nu}$ has been added
to (\ref{4.2}). In the analysis of (\ref{4.1}) the corresponding correction
is also needed, as can be seen in Fig.\thinspace 4.

We test the over-relaxation algorithm by studying the dynamical
scaling behavior\cite{Zheng}. To use this procedure one should start with
a configuration as far as possible from equilibrium with an energy
fixed to its critical value. We take a random configuration for the spins
as a starting configuration. It turns out to have an energy close to zero.
By moving the spins step by step in integrating the differential equation 
(\ref{2.19}) the energy is increased until the critical value is reached.
Then with the of over--relaxation dynamics the energy stays always at
the critical value, but the correlation length should start growing like
$ \xi \propto t^{1/z}$ as a function of the number of update steps $t$.
Finally after many steps the equilibrium situation is reached, that is for
$\xi\approx L$. Therefore the temperature as a function of a moderate number of
updates $t$ should follow a power law relation like (\ref{4.1})
\begin{equation}
T_t = T_L\, + \,T_1\,t^{-1/(z\,\nu)} \ .
\label{4.3}
\end{equation}
using $|T_t - T_L|\, \propto \,t^{-1/(z\,\nu)}$.

The magnetization $|\vec M|$ exhibits more clearly
the power law dependence, since no constant term must be subtracted.
It should depend on the update steps $t$ like
\begin{equation}
\langle | \vec M | \rangle
\propto \,t^{\gamma/(2\,z\,\nu)}\ .
\label{31d}
\end{equation}
This corresponds to $\langle \vec M^2 \rangle \propto L^{\gamma/\nu}$
or $\langle | \vec M | \rangle \propto L^{\gamma/2\,\nu}$, valid for
finite size scaling with the exponent $\gamma$ for the susceptibility.
The rule of transcription is simply that one should replace
$L$ by $t^{1/z}$. Since $\eta$ is small we take 1 for the ratio
$\gamma/(2\,\nu) =1 - \eta/2$.

We used two slightly different ways for the implementation of the
over--relaxation algorithm. The first one consists of
choosing randomly the spins to be updated. The second method uses
a sequential update. The lattice is partitioned into two sublattices
corresponding to the up and down sublattices of an anti--ferromagnet.
Since the interaction is only between nearest neighbors the numbering
of the spin used for the sequential update does not matter. First the
the spin vectors sitting on the up--spin sites are updated and then the
others. Technically one decides whether an index is even or odd before
one starts with the over--relaxation. A spin is an even or odd one if
the sum of its coordinates $(i,\,j,\,k)$ is even or odd.

Taking the results of the simulation depicted in Fig.\thinspace 5 one
notices that with $z_{seq} =1/0.58 \approx 7/4$ the sequential update is the
more efficient algorithm. However, to be more on the safe side concerning
ergodicity one might prefer the random update with
$z_{ran} = 1/0.42 \approx 7/3$. Since always half the total number of spins
are ``turned around'' the greater efficiency of the sequential
update might be connected to the nonlocal nature of the update
reminiscent of the very efficient cluster algorithms.

\section{First-order phase transitions}
The micro-canonical technique can be shown to be most suitable in
simulating first order transitions. A $XY$--spin model having such a
transition is an anti-ferromagnet on the fcc--lattice. For simplicity
we take only the interaction between nearest neighbors.
This model has been analysed by Diep and Kawamura
\cite{Diep89} using standard Monte Carlo techniques. There the
first order nature of the transition for $XY$ and Heisenberg
spins was shown. Especially for an Ising model the frustrated
anti-ferromagnetic order on a fcc--lattice leads to a strong first order
transition \cite{Hu98,Liebm86}. For the vector spins this
phenomenon is less pronounced.

The temperature for a large range of energies has been determined
and the region between 1.35 and 1.55 was scanned more carefully.
The energies are all positive and therefore all temperatures
negative since we kept the convention of the preceding sections
that the ferromagnet has the lowest energy and the exchange constant
$J = -1$. In the inset of Fig.\thinspace 6 the S--shape of the energy
versus temperature dependence clearly indicates the first order
nature of the transition.

We want to analyse in greater detail the data for the largest lattice
with $4\times 24^3$ spins. The idea is to use Maxwell's equal area
construction to determine the transition temperature. A direct approach
is not advisable since the statistical errors for the temperature
changes are quite large. A better way is to determine first the entropy
\cite{Ja98,Hu98}.
With the energy density $\epsilon = E/N$, the entropy density
$s = S/N$, the reciprocal temperature $\beta = 1/T$ and
$ \beta = \partial s/\partial \epsilon$ one obtains the entropy
by integrating
\begin{equation}
\hbox{$1\over N$}\,\ln P = s - \epsilon\,\beta_c =
\int_{\epsilon_0}^{\epsilon} d\epsilon\, (\beta - \beta_c) \ .
\label{5.1}
\end{equation}
The logarithm of the density of states $W$ as the entropy
is changed by subtracting $E/T$, that is $P = W\cdot{\rm e}^{-E/T}$.
This probability $P$ depicted in Fig.\thinspace 7 has the double peak
form of the histograms one obtains by counting the frequency an energy $E$
occurs in a Monte Carlo simulation. The peaks must have equal height at
the transition temperature corresponding to Maxwell's equal area
rule.

The temperature we find by this procedure is $T_c = 0.7397$. However,
in looking at the insert of Fig.\thinspace a considerable one can
observe a consiberable size dependence. The
energy difference between the two maxima $\Delta \epsilon = 0.13$
gives an estimate for the latent heat. One can also read off from
Fig.\thinspace 7 the difference in the logarithm of the probabilities
$\ln P_{max}- \ln P_{min} \approx N*0.0005$, that is for
$\ln P$ at the peaks and for the minimum between the peaks.
Since the cluster has $N = 4*16^3$ spins the ratio
$P_{min}/P_{max} \approx {\rm e}^{-8}$ or $10^{-4}$ is already
too small for a standard Monte Carlo simulation. In upgrading
the weights by using the ``multi-canonical'' method this
low probability problem is circumvented. With the micro-canonical
procedure no modification seems to be necessary here.
However, first order transitions are beset of hysteresis phenomena,
they are bound to occur also in micro-canonical simulations.

\section{Concluding remarks}

We confirmed that the temperature calculation as averages given by
(\ref{2.10}) and (\ref{2.14}) give reliable results. The severe tests were
(a) the size dependence of the temperature for a second order transitions
and (b) the non monotonic dependence of the temperature on the energy in
the neighbour--hood of a first order transition. Further we wanted to
demonstrate that efficient micro-canonical simulations for vector
spins are feasible making use of the over relaxation algorithm.
However, since the histogram technique is missing, the critical energy
cannot be determined precisely by Binder's cumulant technique\cite{Binder}.
For first order transitions the method appears suitable.
 
The inverse of the specific heat as the derivative of the temperature
with respect to energy is linked to averages of fluctuations.
They are far more complicated than the energy fluctuation
one has to calculate for the specific heat in the canonical ensemble.
Nevertheless we could show that the specific heat can be calculated
micro--canonically. That the temperature definition is not unique
is the essential point of Rugh's micro--canonical temperature
definition\cite{Ru98}. This freedom of choice was essential
for finding manageable expressions.

\subsection{Acknowledgment}
 
We thank W.\ Janke for pointing to the close similarity
between a micro canonical and canonical analysis of first order
transitions. We are thankful D.\ Loison for showing  us
the use of the over--relaxation algorithm.

\newpage

\section{Appendix}
\subsection{Micro canonical Specific heat for a harmonic oscillator chain}

The concept used for the micro canonical temperature and the specific heat
determination is best be illustrated for a set of harmonic oscillators.
We take a one dimensional example with the potential energy
\begin{equation}
V = \hbox{$1\over2$} \,
\sum_i^N (x_i - x_{i+1})^2
\quad\hbox{and}\quad x_{N+1} = x_1
\label{A.1}
\end{equation}
for a chain of coupled oscillators.
Following the procedure for the spin chain of section III
a vector $\vec X$ is defined by
\begin{equation}
\vec X = -\Bigl(0,\,F_2,\,0,\,F_4, \ \cdots \>0,\,F_N \Bigr) \Big/
\sum_{i=1}^{N/2} \,F_{2i}^2 \ \,,
\label{A.2}
\end{equation}
where the expressions $F_n = -\partial V / \partial x_n$ mean the forces
acting on the  n-th particle. Obviously Rugh's condition
for the ``flow'' $\vec X$, namely $-\sum_i^N F_i\,X_i = 1$ holds,
since only even components appear in the scalar product. The minus
sign comes from the definition of the forces $F_i$ as negative
derivatives.

The inverse temperature defined as the divergence of $\vec X$ is then
\begin{equation}
{1\over {\cal T}'} \, = \, -
\sum_{i=1}^{N/2} \> \partial_{2i} \>
\> {F_{2i} \over \sum_{j=1}^{N/2} F_{2j}^{\,2} }
\> = \>
{\sum_{i} \partial_{2i}^{\,2}\, V
\over
\sum_j F_{2j}^{\,2} }
\> - \> 2 \,
{\sum_{i} F_{2i}^{\,2} \>
\partial_{2i}^2 V
\over
\bigl[\, \sum_j F_{2j}^{\,2} \bigl]^2}
\label{A.3}
\end{equation}
with $\partial_{2i} = \partial/\partial x_{2i}$. Since the second
derivative of the chain potential is $\partial_{n}^{2} V = 2$
one gets finally
\begin{equation}
{1\over {\cal T}'} \, = \,
\> {N-4 \over \sum_{j=1}^{N/2} F_{2j}^{\,2} } \ ,
\label{A.4}
\end{equation}
that is the expression one has to average to obtain the inverse
temperature.

Neglecting the finite size corrections we determine the temperature
by averaging the quantity
\begin{equation}
{\cal T} \, = \,- {1\over N}\, \sum_{j=1}^{N/2} F_{2j}^2 \ .
\label{A.5}
\end{equation}
The specific heat $C$ is then given by
\begin{eqnarray}
{1 \over C}\, &=& \, { d \over d E }  \> \langle {\cal T} \rangle
\, = \,
\Bigl\langle \sum_{i=1}^{N/2} \> \partial_{2i} \>
\> {F_{2i} \over \sum_{j=1}^{N/2} F_{2j}^{\,2} } \> {\cal T}
\Bigr\rangle
\, + \,
\Bigl\langle {\cal T} \Bigr\rangle \,
\Bigl\langle {1 \over {\cal T}'} \Bigr\rangle
\nonumber \\
\, &=& \,
1  \, + \, \Bigl\langle {\cal T} \Bigr\rangle \,
\Bigl\langle {1 \over {\cal T}'} \Bigr\rangle
\label{A.6}
\end{eqnarray}
Using the expansion (\ref{3.7}) for the reciprocal temperature the last
equation for the inverse of the specific heat reduces to
\begin{equation}
{1 \over C} \, = \,
{4 \over N}  \, - \,
{\langle ({\cal T} - \langle{\cal T} \rangle)^2 \rangle
\over \langle {\cal T} \rangle^2}
\label{A.7}
\end{equation}
in the limit for $N \gg 1$. This is eq.(\ref{3.10}) used for a comparison
with the specific heat formula (\ref{3.9}) of the $XY\!$--spin chain. The
last term of (\ref{A.7}) is the temperature fluctuation. Its size must be
$2/N$ since the specific heat of $N$ oscillators is $C = N/2$. One could
determine this temperature fluctuation for an oscillator chain
analytically, since $V$ and $\cal T$ are quadratic forms, but this
technical detail would be of little interest.

The last formula has the same structure as the specific heat formula of
Lebowitz, Percus and Verlet\cite{Le67} for a system of interacting
particles. There the total conserved energy is the sum of the kinetic and
the potential energy. From the fluctuations of the kinetic energy $K$
(proportional to the fluctuations of the temperature) the specific heat $C$ can
be determined in the micro canonical simulation with the help of
\begin{equation}
{\langle(K - \langle K \rangle)^2\rangle \over \langle K \rangle^2}
\, = \, {2 \over 3\,N} \, - \,
{1 \over C} \ ,
\label{A.8}
\end{equation}
where the average of the kinetic energy for $N$ particles ``measures''
the temperature by $\langle K \rangle = 3 N\,T/2 $.
For the oscillator chain discussed here only the potential energy
was considered (see (A.1)) and the partition of the energy is between
even and odd sites.

\newpage

\begin{figure}
        \begin{center}
                \epsfig{file=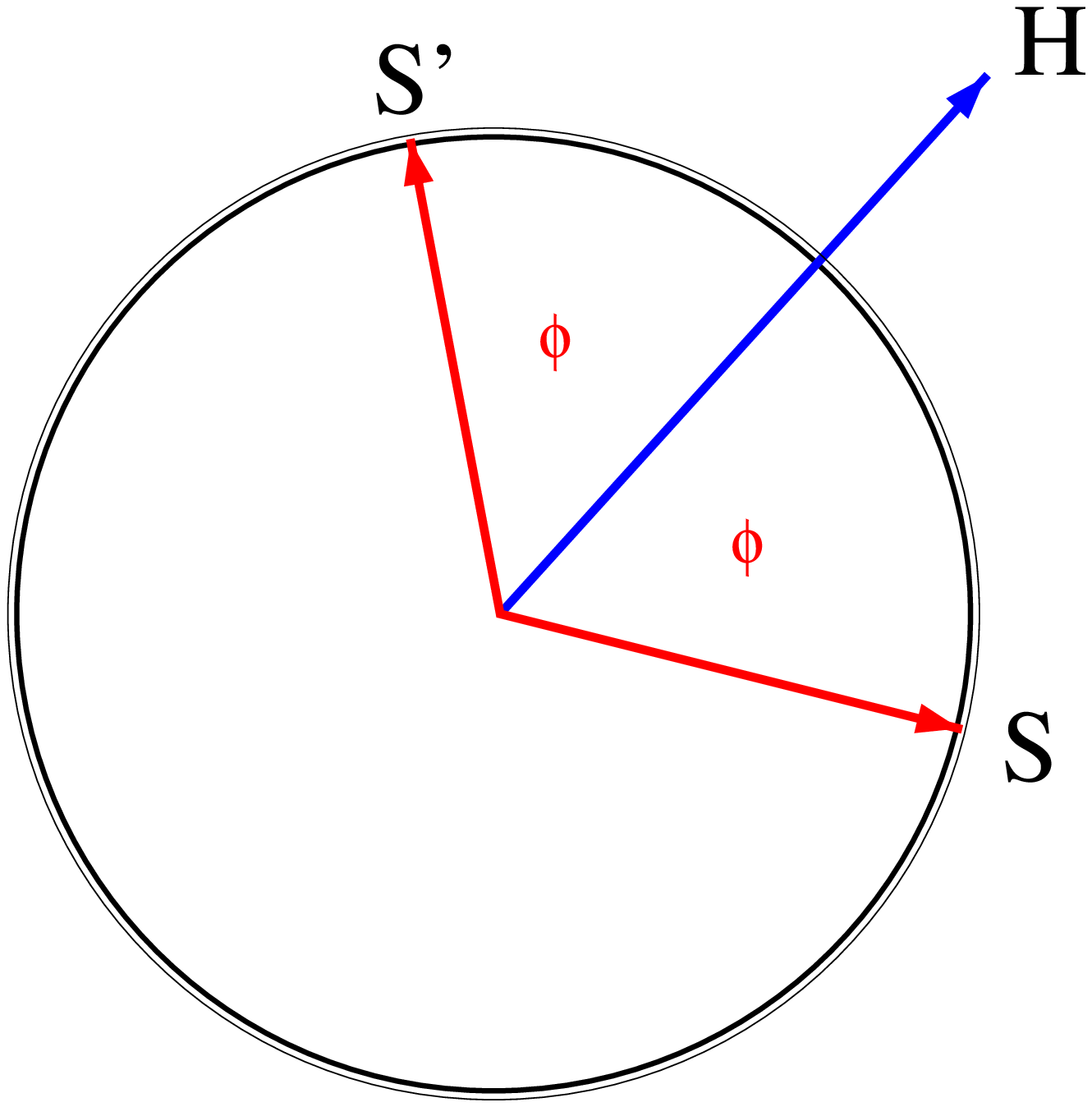,width=3.5in}
        \end{center}
                \caption{Over--relaxation update: 
Mirroring $S$ at the local molecular field 
$H$ produces $S'$.
}
\label{Fig.1}
\end{figure}
\begin{figure}
        \begin{center}
                \epsfig{file=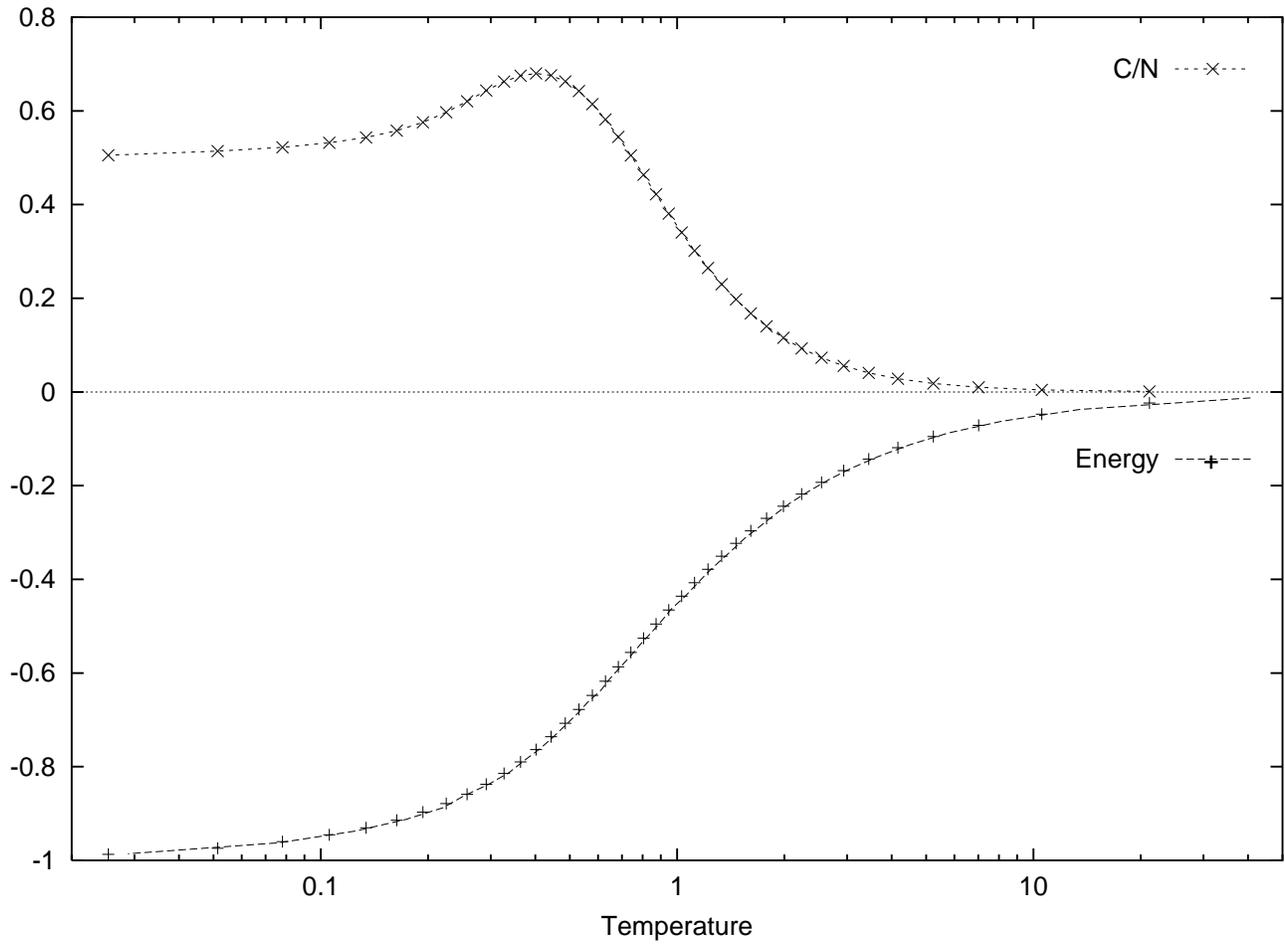,width=5in}
        \end{center}
                \caption{Specific heat and temperature
per spin as a function of temperature for $XY\!$--chain
of 16 spins. The lines represent the canonical relations (14) 
and (15).}
\label{Fig.2}
\end{figure}
\begin{figure}
        \begin{center}
                \epsfig{file=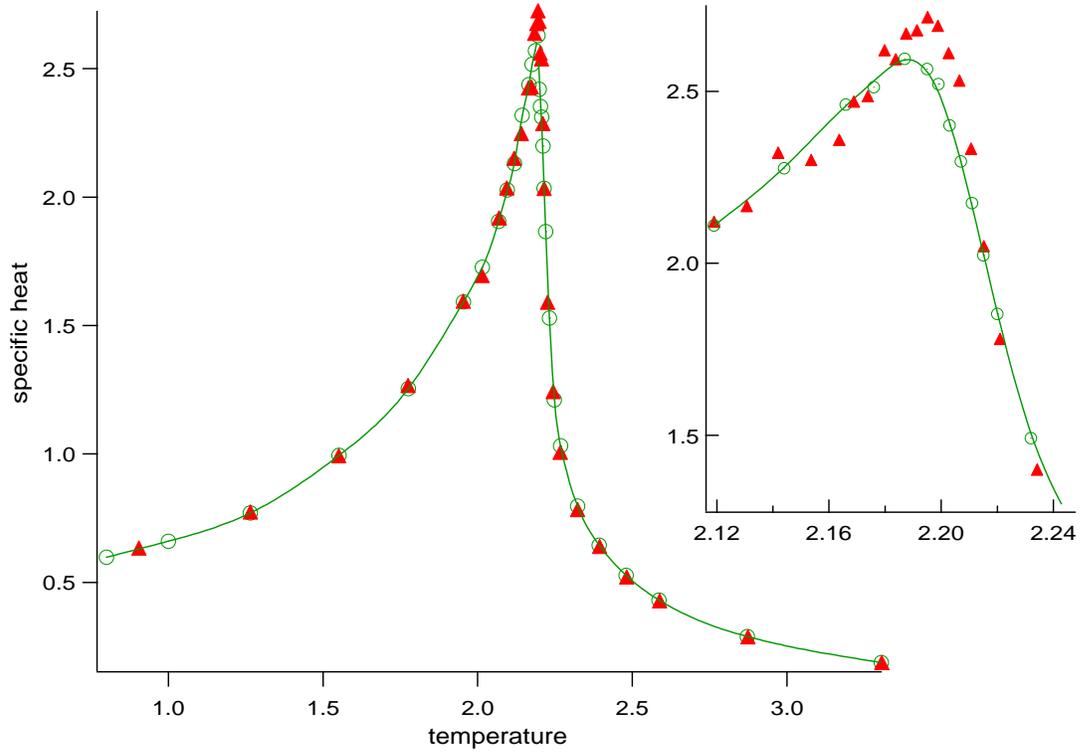,width=6in}
        \end{center}
                \caption{Specific heat of a $24^3$ cubic lattice,
circles canonical and triangles micro--canonical simulations.
Near $T_c \approx 2.20$ canonical ($2 \!\cdot\!10^5$ steps)
and micro--canonical results ($2\!\cdot\!10^6$ steps) differ.
}
\label{Fig.3}
\end{figure}
\begin{figure}
        \begin{center}
                \epsfig{file=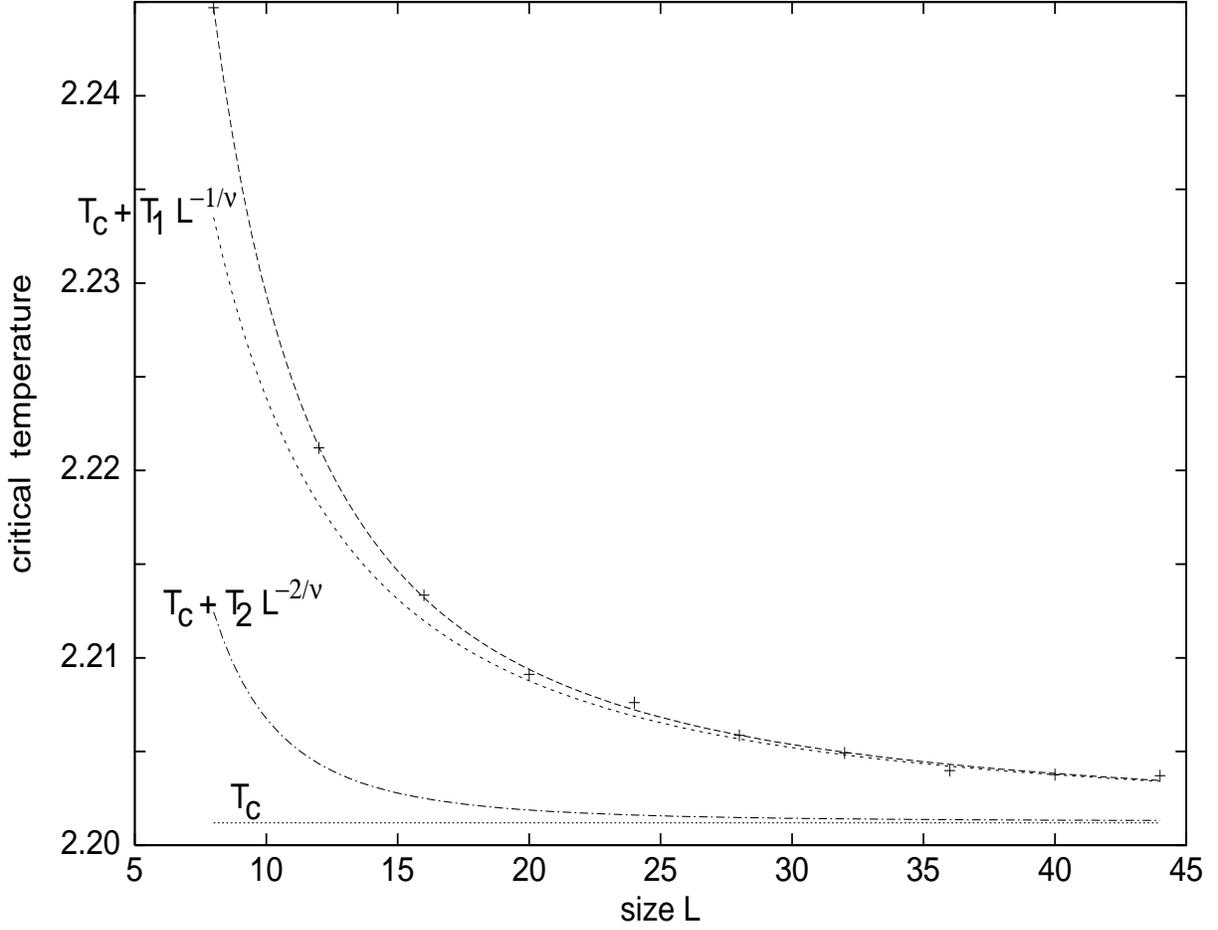,width=5in}
        \end{center}
                \caption{The temperatures ({\rm +}) for
the critical energy $E_c = 0.989$ as a function of the lattice size $L$.
One finds from the fit (upper curve) $T_c = 2.2013$ and $\nu=0.6709$
using eq.(\ref{4.1}) and adding $T_2\,L^{-2/\nu}$
necessary to correct the deviation from scaling for small $L$.
}
\label{Fig.4}
\end{figure}
\begin{figure}
        \begin{center}
                \epsfig{file=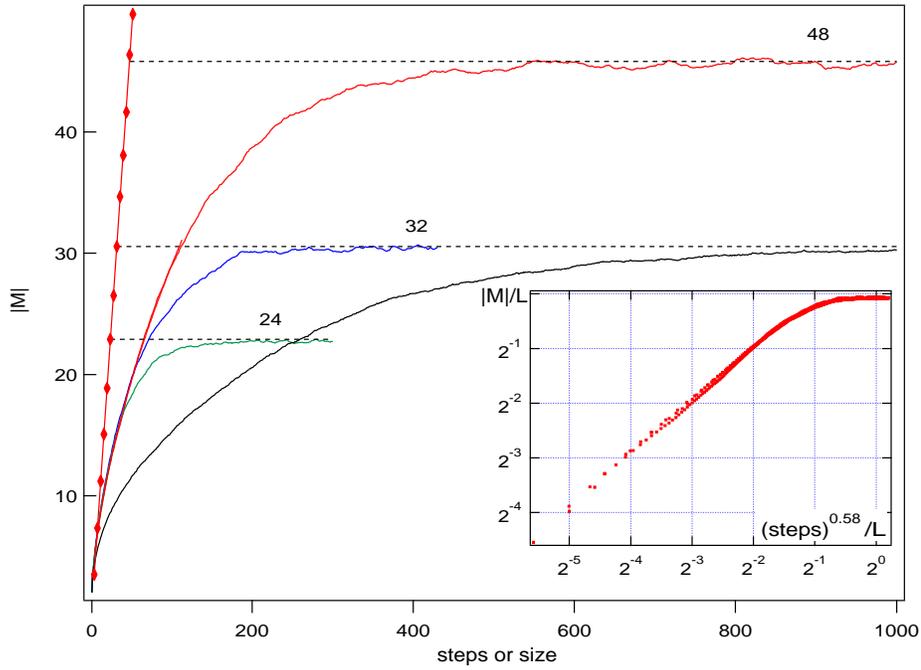,width=5in}
        \end{center}
\medskip
                \caption{
$|M|$ for clusters of size $24^3$, $32^3$ and $48^3$ as a function
of sequential updates starting from disorder at the critical energy.
Also shown is the linear size dependence of $|M|$ used for the
scaling plot in the insert. Not shown is the update with randomly
chosen spins for $32^3$ with an almost linear relation between $|M|/L$
and (steps)$^{0.42}/L$. $|M|$ has been
averaged a few thousand times.
}
\label{Fig.5}
\end{figure}
\begin{figure}
        \begin{center}
                \epsfig{file=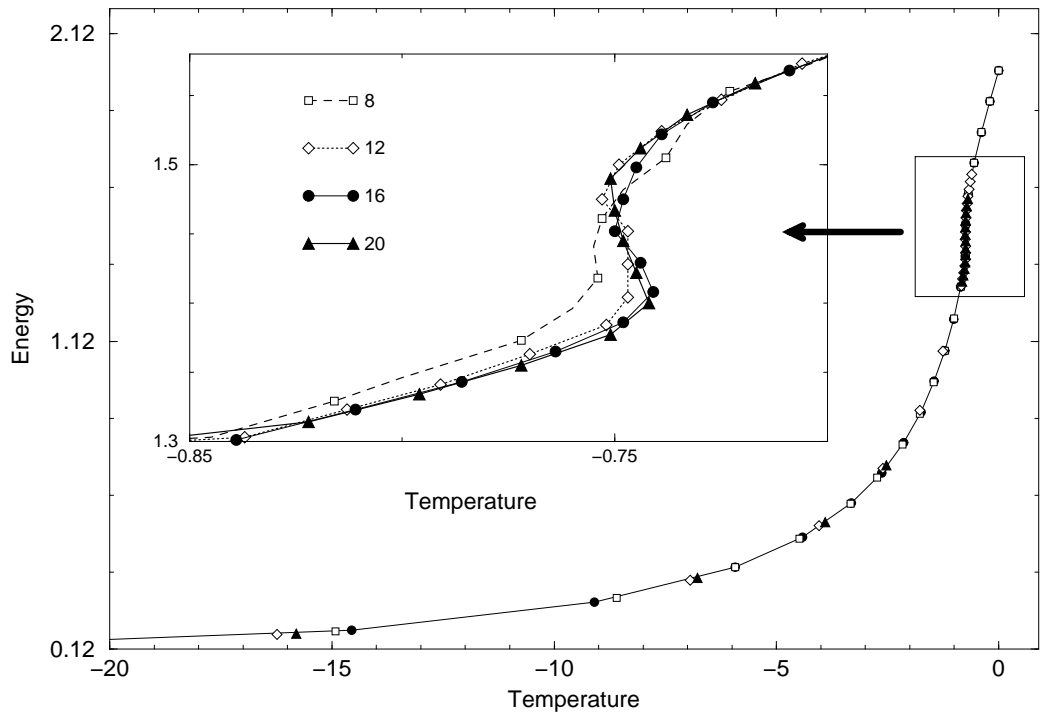,width=5in}
        \end{center}
                \caption{Results of microcanonical simulations for
fcc lattices with linear dimension 8, 12, 16 and 20. The energy
is plotted as a function of the temperature. The inset shows
the S-form of the energy in the vicinity of $T_c \approx 0.75$.}
\label{Fig.6}
\end{figure}
\begin{figure}
        \begin{center}
                \epsfig{file=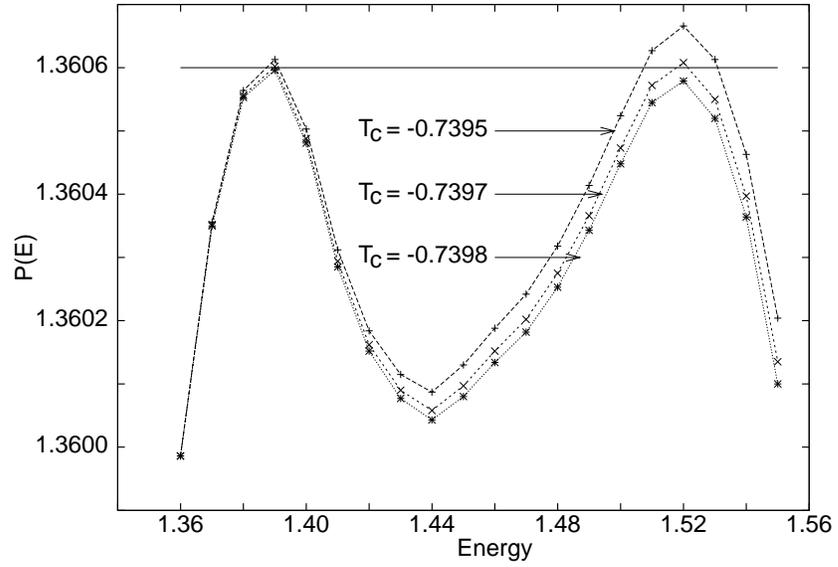,width=3in}
        \end{center}
        \caption{Relation between energy density $\epsilon$ 
and probability
$P(\epsilon) = e^{s(\epsilon)-\epsilon\beta(c)}$ for the antiferomagnet
on a fcc lattice of size $16^3$ according to (\ref{5.1}).
Equal height of the two peaks selects the best critical temperature.
}
\label{Fig.7}
\end{figure}
\end{document}